# Simulation of proton radiography terminal at IMP

YAN Yan(严岩)[1]  SHENG Li-Na(盛丽娜)[2,1)]  HUANG Zhi-Wu(黄智武)[1]  WANG Jie(王洁)[1]

YAO Ze-En(姚泽恩)[1,2)]  WANG Jun-Run(王俊润)[1]  WEI Zheng(韦峥)[1]  YANG Jian-Cheng(杨建成)[2]  YUAN You-Jin(原有进)[2]

[1]School of Nuclear Science and Technology, Lanzhou University, Lanzhou 730000, China

[2]Institute of Modern Physics, Chinese Academy of Sciences, Lanzhou 730000, China

**Abstract:** Proton radiography is used for advanced hydrotesting as a new type radiography technology due to its powerful penetration capability and high detection efficiency. A new proton radiography terminal will be developed to radiograph static samples at Institute of Modern Physics of Chinese Academy of Science (IMP-CAS). The proton beam with the maximum energy of 2.6 GeV will be produced by Heavy Ion Research Facility in Lanzhou-Cooling Storage Ring (HIRFL-CSR). The proton radiography terminal consists of the matching magnetic lens and the Zumbro lens system. In this paper, the design scheme and all optic parameters of this beam terminal for 2.6GeV proton energy are presented by simulating the beam optics using WINAGILE code. My-BOC code is used to test the particle tracking of proton radiography beam line. Geant4 code and G4beamline code are used for simulating the proton radiography system. The results show that the transmission efficiency of proton without target is 100%, and the effect of secondary particles can be neglected. In order to test this proton radiography system, the proton images for an aluminum plate sample with two rectangular orifices and a step brass plate sample are respectively simulated using Geant4 code. The results show that the best spatial resolution is about 36μm, and the differences of the thickness are not greater than 10%.

## 1. Introduction

For more than half a century, the main tool for hydrodynamic experiment was pulsed x-ray radiography. It allows one to see the inside of an object with a complex structure without disturbing it. However, the dose limitations, position resolution and backgrounds still limit the utility of x-ray radiography for obtaining the pictures with higher resolution. About 50 years ago, proton radiography technology, which is predicted to achieve higher spatial resolution, was put forward. However, the experimental results of earlier research show that proton radiography has a lower spatial resolution comparing with X-ray radiography[1-3]. The main reason is that the interaction between the protons and the sample atoms, especially multiple Coulomb scattering (MCS) lead to the image blur. The problem hasn't been solved until Mottershead and Zumbro[4] proposed a typical magnetic quadrupole lens system for the correction of image blur in 1997. The proton radiography system has been successfully developed at some laboratories, such as LANSCE[5-7] in US, ITEP and IHEP in Russia[8, 9] and GSI in Germany[10]. In addition, some new proton radiography systems are being developed at FAIR in Germany[11] and IMP in China[12]. The IMP-CAS has proposed two proton radiography

*Supported by the NSAF Joint Funds of National Natural Science Foundation of China (Grant No. 11176001).

1) E-mail: shenglina@impcas.ac.cn

2) E-mail: zeyao@lzu.edu.cn

beam line[13] to investigate proton radiography technique. This paper is mainly to develop the proton radiography terminal of IMP, including calculation of the beam optic parameters and the simulation of proton radiography by Monte Carlo method.

## 2. The Zumbro lens system

The attenuation of a proton beam through a given density of material is described by the equation:[5]

$$N/N_0 = e^{-l/\lambda} \qquad (1)$$

where $N_0$ and $N$ are respectively the incident number and the transmitted number of proton, and $l$ is the path length of the object, $\lambda$ is the mean free path of proton. The transmitted protons undergo multiple coulomb scattering (MCS) with the atom of target so it will produce a angular distribution after the target. This angular distribution can be approximately described by Gauss distribution, the half-width of this Gauss distribution is expressed as[14]

$$\phi_c = \frac{0.0136 GeV}{\beta cp}\sqrt{\frac{z}{X_0}}\left[1+0.038 In\left(\frac{z}{X_0}\right)\right] \qquad (2)$$

where, p is the beam momentum, $\beta c$ is the velocity of the proton, z is the thickness of the object, $X_0$ is proton's radiation length for material[15], its value is related to the material's mass number and atomic number.

The angle distribution produced by MCS will lead to the image blur seriously. So in order to cancel the effect by MCS, a typical lens was designed by Mottershead and Zumbro which is called Zumbro lens. The Zumbro lens system[4] is the type of angle-matching lens which is widely used to focus the particles to achieve point-to-point imaging for removing most image blur by MCS. Furthermore, the magnetic lens system can provide angle sorting which allows to insert an angle cut aperture to aid material identification.

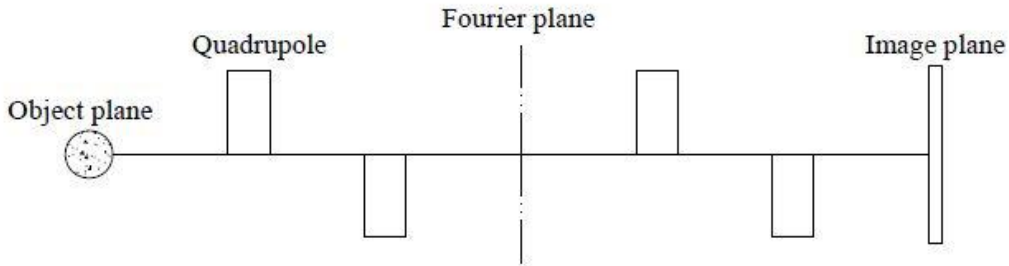

Fig. 1 The schematic of the Zumbro lens

Zumbro lens system consists of two identical quadrupole doublet cells, each cell includes a positive quadrupole and a negative quadrupole, as shown in Fig. 1. This quadrupole beam line is reflection symmetric structure because its second cell's parameters are identical with the first one. The reflection symmetric lens focuses the particles at image plane with the magnification of 1 along x and y axises. In order to achieve above aim, the lens transfer matrix should be $-\vec{I}$. If the transfer matrix of one cell is M, that the whole lens system is given by the 2×2 matrix identity

$$\vec{R} = \vec{M}^2 = -\vec{I}\det(\vec{M}) + \tau\vec{M} \tag{3}$$

where $\tau \equiv Tr(\vec{M}) = M_{11} + M_{22}$ is the trace of $\vec{M}$, let $\tau = 0$, the determinant $\det\vec{M} = 1$ for beam line matrices. So the lens transfer matrix $\vec{R} = -\vec{I}$, which achieves an inverse image with the same size. Considering the particle's momentum deviation $\Delta = \delta p/p$, the final beam spot size of this particle at the image plane is (to first order in $\Delta$)

$$x_f = R_{11}x + R_{12}\theta + (R_{11}' + \omega R_{12}')x\Delta + R_{12}'\varphi\Delta \tag{4}$$

where $(x,\theta)$ is the particle's initial coordinate, $\omega = L^{-1}$ is the correlation coefficient, L is the distance between particle source and the entry plane of the lens. A beam would lie along the line $\theta = \omega x$ in phase space, here if we make

$$R_{11}' + \omega R_{12}' = 0, \text{ or } \omega = -R_{11}'/R_{12}' \tag{5}$$

all position dependent chromatic aberrations vanish, so the angle sorting function can be carried out. With $R_{12} = 0$, $R_{11} = -1$, the final position is

$$x_f = -x + R_{12}'\varphi\Delta \tag{6}$$

The remaining chromatic aberration depends only on the deviation angle $\varphi$.

### 3. Design of the proton radiography beam line

Before design the proton radiography beam line, the appropriate energy of proton should be decided first. If only considering the penetrating power of proton, the 1GeV proton is enough for the target. But the spatial resolution must also be considered. The main factor is the MCS effect, which will reduce the spatial resolution, the value of position error can be estimated by [16]

$$\Delta r = \theta_0 l/\sqrt{3} \tag{7}$$

where $\theta_0$ is RMS value of scattering angle by MCS, $l$ is the thickness of target. In equation (7), $\Delta r \propto \theta_0$, $\theta_0$ can be derived by

$$\theta_0 \approx \frac{14.1}{p\beta}\sqrt{\sum_i^n \frac{\rho_i l_i}{R_i}} \tag{8}$$

where p is the momentum of proton, $\beta = v/c$ is the relative velocity of proton, $\rho_i$ and $l_i$ are the density and thickness of target respectively, $R_i$ is the radiation length of proton in target material. For a determined target, $\rho_i, l_i$ and $R_i$ are constants. So $\theta_0 \propto 1/p\beta$. According to equations (7) and (8), we can derive $\Delta r \propto 1/p\beta$, That is to say the choice of higher energy protons can effectively reduce $\Delta r$ value. The maximum design energy of proton is 2.6 GeV in HIRFL-CSR, so in order to reduce $\Delta r$ we choose 2.6 GeV in this paper to simulate the proton radiography.

According to the beam condition of HIRFL-CSR at IMP, 2.6 GeV proton radiography terminal is designed, which consists of a matching section, an imaging lens system and proton detector. Fig. 2 shows the schematic of 2.6 GeV proton radiography beam line scenario. Q1, Q2 and Q3 are the matching lens system, and Q4, Q5, Q6 and Q7 are the imaging lens system.

The matching lens section consists of 3 quadrupole lens to provide the required phase space correlation upstream of the object. Moreover, this section also expands the beam's transverse size to illuminate the field of view of the object fully. The Zumbro magnets structure is used here in imaging lens system, which will eliminate a major part of the chromatic and geometric aberrations, and match the rays with different MCS angle. Moreover, a collimator at the Fourier plane is used to eliminate the larger angle scattered particle to enhance the spatial resolution and the capability of material identification.

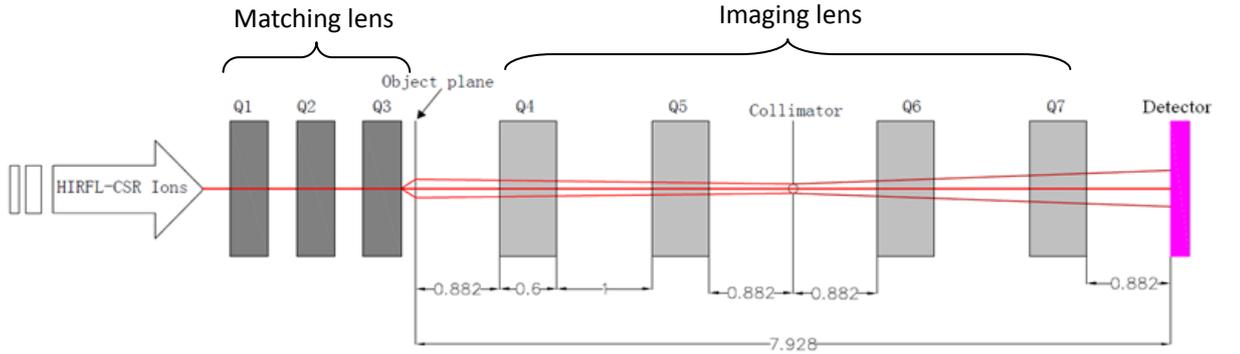

Fig. 2 The schematic of 2.6GeV proton radiography beam line scenario (unit/m)

According to adjust the parameters from equation (4)-(6), make the imaging lens system meets the Zumbro magnets requirements, and the magnifications are both 1 along x and y axises. In order to meet above requirements, the corresponding parameters for the 2.6GeV proton radiography beam line are calculated. The gradients and geometrical parameters of the imaging lens are calculated by WINAGILE code[17], Fig. 3 shows the beam envelop of the beam line by WINAGILE. The values of the parameters are listed in Table 1. With the parameters in Table 1, the particle transport trajectory in the imaging lens is simulated by My-BOC code[18], it can be seen the proton trajectory in image lens achieve the point-to-point imaging in Fig. 4.

Table 1. Parameters of the imaging lens system of 2.6GeV proton radiography beam line.

| Parameters | Values |
|---|---|
| Proton energy(GeV) | 2.6 |
| Magnet aperture(mm) | 100 |
| Field gradient of imaging lens(T/m) | 15.53(Q4,Q6),-15.53(Q5,Q7) |
| Total length(m) | 7.9 |
| Field-of -view(mm) | 20 |
| Magnification x/y | 1/1 |
| Beam emittance(πmmmrad)x/y | 10/10 |
| Beam parameters in object plane x/y | 4.524/-4.524(Alpha) |
| | 10/10(Beta) |
| Beam parameters in image plane x/y | 4.524/-4.524(Alpha) |
| | 10/10(Beta) |

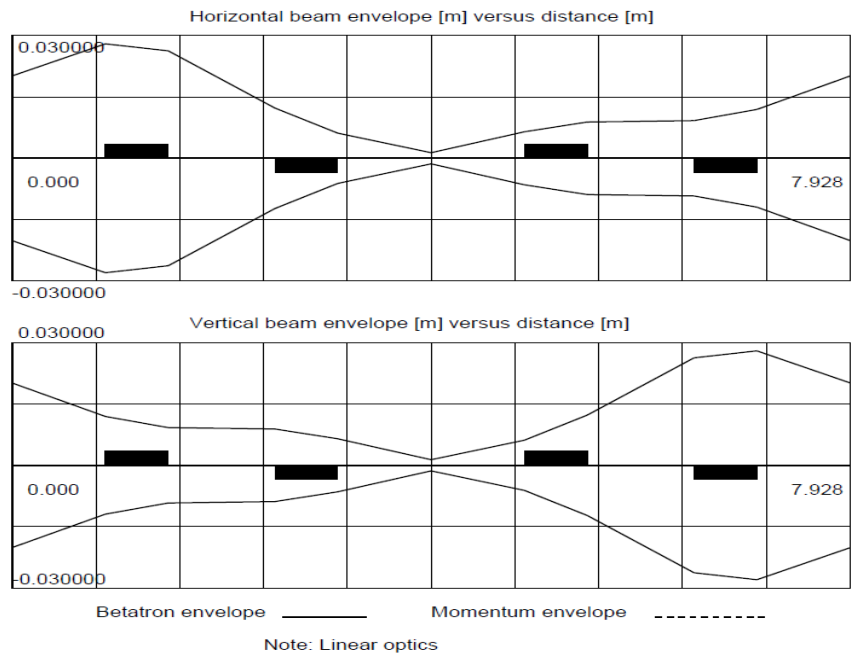

Fig. 3 Beam envelope of the beam line

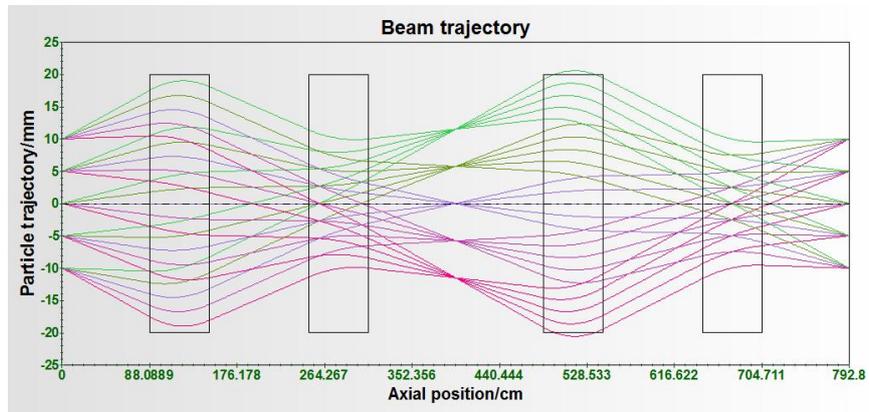

Fig. 4 The particle tracking of proton radiography

## 4. Monte Carlo simulation

### 4.1 Monte Carlo model

In order to test the designed proton radiography system (see Fig. 2), the beam parameter and the image were respectively simulated using Monte Carlo codes Geant4[19] and G4beamline[20]. The parameters of the proton radiography system were setup according the schematic in Fig. 2 and the data in Table 1. During the simulation process, the class of QGSP_BIC in Geant4 is invoked to achieve all essential interactions, including Coulomb interaction between the protons and the electrons, strong interaction between the protons and the atomic nuclei. Some self-field effects were ignored, such as the space charge effect and fringe field effect. The ideal detectors are respectively located at the Fourier plane and the image plane to obtain the beam parameters and image results.

**4.2　The simulation results of the beam parameters**

Based on above model, the protons angle distribution is firstly simulation using Geant4 code. Here the incident number of proton is $1\times 10^6$. The simulated results are compared with the computed data by equation (1) and (2), they agree each other. So self-field can be ignored, the simulation code can be trusted.

The beam parameters are also investigated using G4beamline code. The results are shown in Fig. 5, the phase space diagram at the object plane (Fig. 5(a)) and the image plane (Fig. 5(c)) are identical each other in horizontal and vertical directions. It means when proton pass though the imaging lens, the magnification of proton at image plane is 1. And the horizontal and vertical phase spaces are superposition at the Fourier plane in Fig. 5(b). The parameters of the beam are totally the same as the calculated results, meet the design requirement.

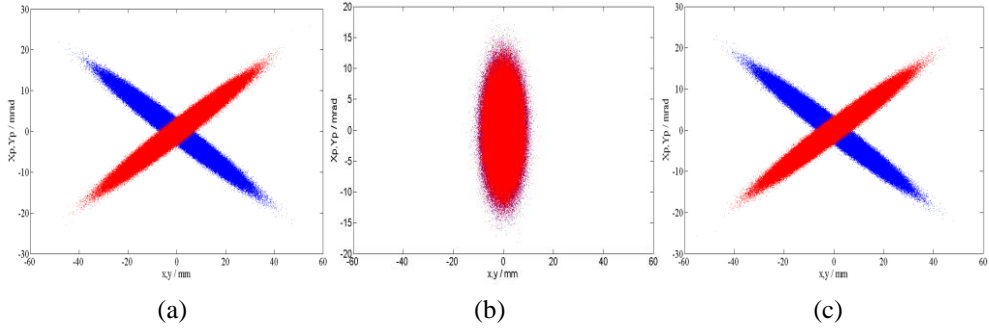

(a)　　　　　　　　　　　(b)　　　　　　　　　　　(c)

Fig. 5 Horizontal and vertical phase space diagram (the blue part and red part are the phase space in horizontal and vertical directions respectively) (a) at the object plane (b) at the Fourier plane (c) at the image plane

In order to get the high resolution image, high transportation efficiency of the beam line for 2.6 GeV protons must be ensured from particle source to image plane without target. So we input the optic parameters into G4beamline to simulate the beam line without target to check the transmission efficiency. In the simulation, the number of incident particles is $10^6$. An ideal detector is located at the image plane to collect protons, and the results show all incident protons are collected by the ideal detector, it means the transmission efficiency is 100%, this result by Monte Carlo code is agree with the result using WINAGILE code and My-BOC code.

When the protons pass through the target, a lot of secondary particles, such as gamma, neutron, proton, electron, $\pi$ meson, k meson, will be produced. These secondary particles may influence the image quality. So the secondary particles need to be investigated. The secondary particles transportation in the proton beam line is simulated using Geant4 code. The results show that only 11 particles arrive at the image detector with a 100 mm diameter under $10^5$ incident protons. A reasonable explanation is that almost all of the charged particles are unable to reach the image plane for these particles cannot pass through the focus lens, and a majority of neutral particles are also lost in the beam pipe. Thus, the effect of the secondary particles can be neglected.

### 4.3 The simulation results of the sample image

In order to test the image effect of the designed proton radiography beam line, the images of two samples are simulated using Geant4 code. The first sample for testing the spatial resolution is a circular aluminum object (2 cm thickness) with two 10 mm×3 mm rectangular orifices in the center. Another sample for testing the thick resolution is a step brass plate with three layers to test the thick resolution. Fig. 6 shows the structures and geometry sizes of two samples.

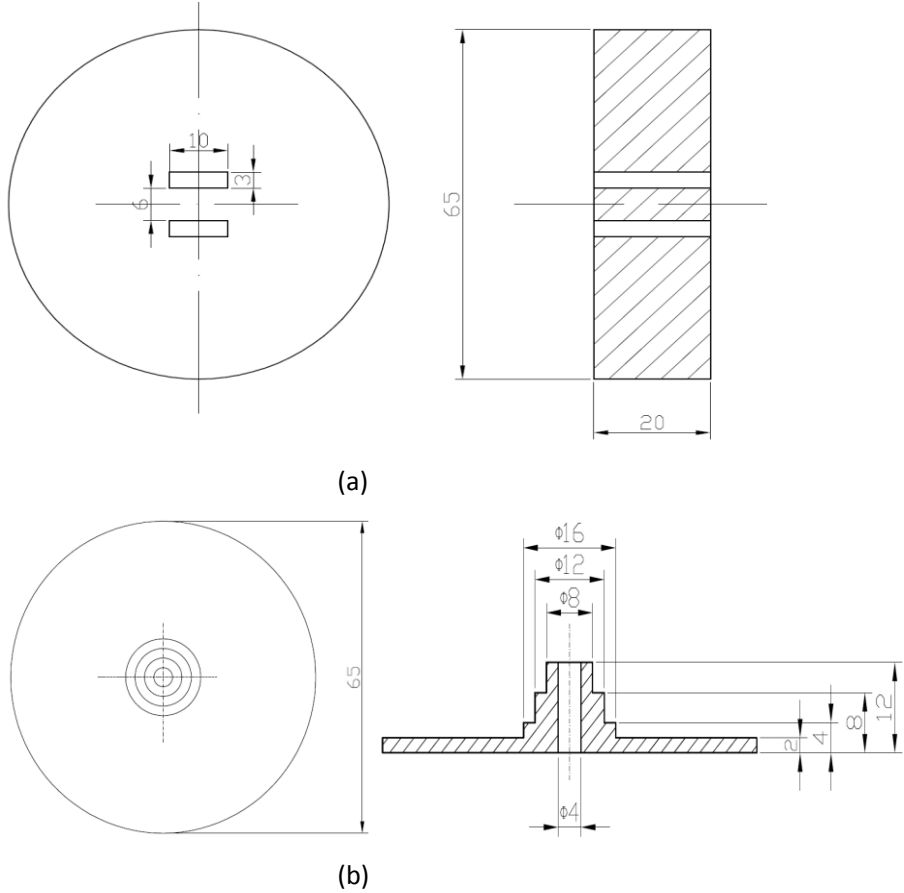

(a)

(b)

Fig. 6  (a) Drawing of orifices sample (unit/mm) (b) Drawing of steps sample (unit/mm)

Fig. 7(a) shows the two dimensional image of the first sample on x-y plane, the two rectangular orifices can be seen clearly. Fig. 7(b) is the proton counts distribution along the x axis, the two steps in the distribution curve are corresponding to the position of the rectangular orifices. Fig. 7(c) is obtained by a differential operation for the data in Fig. 7b. Four peaks marked as 1,2,3,4 can be clearly seen in Fig. 7(c). The Sigma of four peaks, which is the standard deviation of the Gauss fitting function, is defined as the position resolution corresponding to four edgy of two rectangular orifices. The data analysis shows that the best resolution is about 36 μm for the edge of rectangular orifices.

According to the same method, the simulation results for the steps sample are obtained, and the results are shown in Fig. 8. Fig. 8 (a) is the two dimensional image of the steps sample on x-y plane. Every step edge can be seen clearly in the image. The same analysis process also be made for the data of Fig. 8(b). In Fig. 8(c), four peaks marked by 1,2,3,4 are the positions of step edge. The spatial resolution between the hole at the center and brass is 83 μm. And the best spatial resolution between different thicknesses of brass steps is about 161 μm.

Moreover, the further simulations are operated to estimate the thickness of brass steps by the imaging data in Fig. 8(b). Brass plates with the thickness of 2mm,4mm,8mm and 12mm are respectively put into the object plane of the beam line as the target to get the transmission ratio of proton. The simulation results are showed in the second column in Table 2. According to the simulation results, a fitting formula between the transmission ratio of proton and the thickness is established:

$$f(x) = p_1 x^2 - p_2 x + p_3 \tag{9}$$

where $p_1$=17.39, $p_2$=-35.38, $p_3$=19.49 are fitting parameters, x is the transmission ratio of proton, f(x) is the areal density of brass target, which is corresponding to the target thickness. The transmission ratio of steps sample are obtained by calculating the average number of proton each steps in Fig. 8(b), and are showed in the third column in Table 2. The calculated thicknesses of the steps sample are obtained by a combination of the third data in Table 2 and equation (9). The maximum differences is about 7.3% between the real thickness and the calculated thickness.

Table 2 Calculation results of step thickness

| Thickness of brass (mm) | Transmission ratio | Transmission ratio of steps sample | Calculated thickness (mm) | Difference with actual value (%) |
| --- | --- | --- | --- | --- |
| 2 | 0.88 | 0.80 | 2.03 | 1.5 |
| 4 | 0.68 | 0.65 | 4.29 | 7.3 |
| 8 | 0.44 | 0.46 | 7.70 | 3.7 |
| 12 | 0.29 | 0.28 | 12.2 | 1.7 |

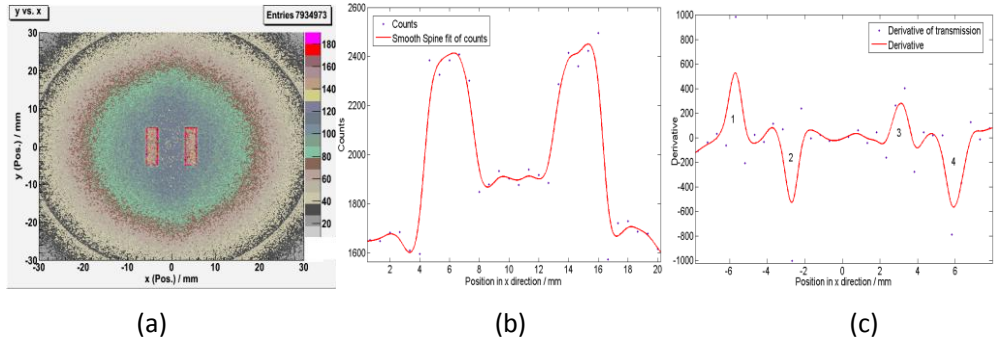

(a)      (b)      (c)

Fig. 7 The simulation results with orifices sample by Geant4 code (a) Image on x-y plane; (b) Transmission in the edge of x direction; (c) Gauss fit for derivative of edge transmission.

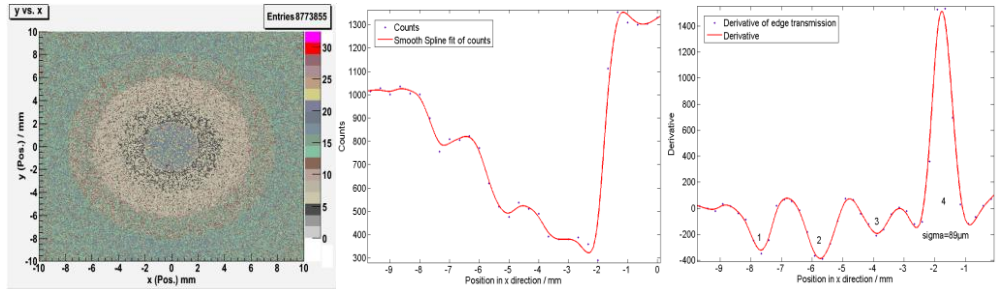

Fig. 8 The simulation results with step sample by Geant4 code. (a) Image on x-y plane; (b) Transmission in the edge of x direction; (c) Gauss fit for derivative of edge transmission.

## 5 Conclusion

A 2.6 GeV proton beam line for proton radiography, which consists of the matching lens and the Zumbro lens system, has been designed. The optic parameters of the matching magnetic lens and the Zumbro lens system are presented by simulating the proton transportation using WINAGILE code and My-BOC code. The transmission efficiency and secondary particles are respectively investigated by

Monte Carlo simulation using Geant4 code and G4beamline code. The results show that the transmission efficiency of proton is 100%, the effect of the secondary particles can be neglected. The simulation test of the proton images for the orifices sample and the steps sample have been carried out using Geant4 code. The structure of two samples can be identified clearly in the images and the best spatial resolution is 36 μm. The difference of the thickness is not greater than 10%.

Considering the actually experiment, the detect efficiency will influence the final resolution, so the simulation for detector is the next step work. Furthermore, the collimator in Fourier plane will be studied, including its aperture and material, which will improve the spatial resolution of the terminal. Moreover, further simulation will be made with higher energy protons in High Intensity heavy-ion Accelerator Facility (HIAF) proposed by IMP, higher spatial resolution can be expected due to higher energy of proton. In addition, the experiment will be carried out at IMP in future.